\documentclass[11pt]{article}

\usepackage[a4paper,margin=1in]{geometry}

\usepackage{amsmath,amssymb,amsfonts,amsthm,mathtools}

\usepackage{graphicx}
\usepackage{booktabs}
\usepackage{array}
\usepackage{caption}
\usepackage{subcaption}

\usepackage{algorithm}
\usepackage{algorithmic}

\usepackage{textcomp}
\usepackage{tikz}
\usepackage{xcolor}
\usepackage{microtype}
\usepackage{enumitem}

\usepackage[numbers,sort&compress]{natbib}
\usepackage[hidelinks]{hyperref}
\usepackage[nameinlink,noabbrev]{cleveref}

\title{\textbf{Beyond TVLA: Anderson--Darling Leakage Assessment for Neural Network Side-Channel Leakage Detection}}

\author{
Ján Mikulec\thanks{Faculty of Informatics and Information Technologies, Slovak University of Technology in Bratislava, Slovakia. Email: jan.mikulec@stuba.sk}
\and
Jakub Breier\thanks{TTControl GmbH, Vienna, Austria. Email: jbreier@jbreier.com}
\and
Xiaolu Hou\thanks{Faculty of Informatics and Information Technologies, Slovak University of Technology in Bratislava, Slovakia. Email: xiaolu.hou@stuba.sk. This work was supported by the Open Research Fund of The State Key Laboratory of Blockchain and Data Security (Grant number: A2566), Zhejiang University}
}

\date{}

\begin{document}

\maketitle

\begin{abstract}
Test Vector Leakage Assessment (TVLA) is widely used to detect side-channel leakage, but its reliance on Welch's \(t\)-test makes it primarily sensitive to differences in the mean of two leakage populations.
This can limit its ability to identify leakage that manifests through broader distributional changes.
We introduce Anderson--Darling Leakage Assessment (ADLA), a distribution-sensitive leakage assessment methodology based on the two-sample Anderson--Darling test.
To enable direct comparison with conventional TVLA, we derive an ADLA decision threshold corresponding to the nominal significance level associated with the standard TVLA threshold of \(4.5\).

We evaluate ADLA on a shuffling-protected embedded multilayer perceptron implementation under fixed-versus-fixed and fixed-versus-random input configurations.
Across the considered settings, ADLA exhibits clearer threshold exceedances than TVLA and identifies leakage locations that are not detected by the mean-based test.
To assess the practical relevance of these locations, we further perform correlation power analysis using points of interest selected from the ADLA and TVLA statistics.
The ADLA-selected points enable recovery of the exponent byte of the targeted model weight in the presence of shuffling.
\end{abstract}

\noindent\textbf{Keywords:} Neural networks; side-channel attacks; leakage assessment; TVLA; Anderson--Darling

\section{Introduction}
Neural networks are increasingly deployed on embedded and edge platforms for applications ranging from vision and biometrics to industrial monitoring and automotive control.
While such deployments enable low-latency inference close to the data source, they also expose neural network implementations to physical attacks~\cite{batina2022implementation,hou2021physical,batina2019csi}.
In particular, side-channel analysis (SCA) attacks~\cite{batina2019csi} can exploit data-dependent variations in physical observables, such as power consumption or electromagnetic emanations, to infer information about intermediate computations, model parameters, or user inputs.
To mitigate such leakage, countermeasures including masking~\cite{dubey2022modulonet}, shuffling~\cite{puvskavc2025make}, and jitter-based desynchronization~\cite{breier2023desynchronization} have been proposed for neural network implementations on constrained devices.

A widely adopted first step in evaluating side-channel resistance is \emph{leakage assessment}, which aims to determine whether statistically detectable data-dependent leakage is present without requiring a specific attack strategy~\cite{isoiec17825}.
The de facto standard methodology is Test Vector Leakage Assessment (TVLA)~\cite{gilbert2011testing,jayasena2023tvla}.
TVLA is attractive due to its simplicity and well-established thresholding practice; however, it is fundamentally a mean-based test.
In protected implementations, leakage may manifest through distributional differences that are not fully characterized by changes in the mean.
This motivates complementary leakage-assessment techniques that can detect distributional discrepancies beyond differences in the mean.

In this work, we propose \emph{Anderson--Darling Leakage Assessment (ADLA)}, a leakage assessment methodology based on the two-sample Anderson--Darling test.
In contrast to TVLA, which tests for differences in the population means, ADLA evaluates whether two leakage populations follow the same distribution, thereby providing sensitivity to a broader class of distributional differences.
We further derive an ADLA decision threshold corresponding to the nominal significance level associated with the conventional TVLA threshold of \(4.5\).

We evaluate ADLA on a shuffling-protected embedded MLP implementation.
Across the evaluated settings, ADLA produces clearer threshold exceedances than TVLA, including leakage locations that remain below the conventional TVLA threshold.
To assess the practical relevance of these locations, we further perform CPA using points of interest selected separately from the ADLA and TVLA statistics.
The ADLA-selected points enable recovery of the exponent byte of the targeted model weight in the presence of shuffling.

Our contributions are summarized as follows:
\begin{itemize}
    \item We introduce ADLA, a distribution-sensitive leakage assessment methodology for neural network implementations based on the two-sample Anderson--Darling test.

    \item We derive an ADLA decision threshold corresponding to the nominal significance level associated with the conventional TVLA threshold, enabling a statistically consistent comparison between the two methods.

    \item We experimentally show that ADLA exhibits stronger leakage-detection responses than TVLA across the evaluated settings.

    \item Using two points of interest identified by ADLA, we recover the exponent byte of a target model weight in the presence of a shuffling-based countermeasure.
\end{itemize}

The remainder of this paper is organized as follows.
Section~\ref{sec:related} reviews the relevant background on neural networks, side-channel attacks, leakage-assessment methodologies, and countermeasures.
Section~\ref{sec:preliminaries} introduces the statistical foundations of TVLA and ADLA.
Section~\ref{sec:ADLA} derives the ADLA decision threshold.
Section~\ref{sec:evaluation} describes the experimental setup and presents the leakage-assessment results.
Section~\ref{sec:cpa} evaluates the practical relevance of the detected leakage through CPA using points of interest selected by ADLA and TVLA.
Section~\ref{sec:discussion} discusses the methodological implications and compares ADLA with alternative leakage-assessment approaches.
Finally, Section~\ref{sec:conclusion} concludes the paper.

\section{Background and Related Work}
\label{sec:related}
This section reviews the background and related work relevant to side-channel leakage assessment of neural network implementations.
We briefly introduce the neural network computations relevant to our evaluation (Subsection~\ref{sec:nn}), review SCA attacks on neural network implementations (Subsection~\ref{sec:scaonnn}), summarize existing leakage-assessment methodologies (Subsection~\ref{sec:la}), and discuss corresponding countermeasures (Subsection~\ref{sec:countermeasure}).

\subsection{Neural Networks}
\label{sec:nn}
Neural networks~\cite{goodfellow2016deep} are computational models composed of layers of interconnected neurons whose computations are governed by trainable parameters, typically weights and biases.
During inference, these parameters, together with the chosen activation functions, determine the linear and nonlinear transformations that produce the network output.

A multilayer perceptron (MLP) is a feedforward neural network composed of an input layer, one or more fully connected hidden layers, and an output layer.
During inference, each neuron computes a weighted sum of its inputs, adds a bias term, and applies a nonlinear activation function. 
For the side-channel setting considered in this work, we focus on inference and, in particular, on the multiply-and-accumulate computations involving input values and model weights.
The specific MLP architecture evaluated in this work is described in Section~\ref{sec:experimental_setup}.

\subsection{Side-Channel Analysis Attacks on Neural Networks}
\label{sec:scaonnn}
SCA attacks on neural network implementations can target different forms of proprietary or sensitive information, including the model architecture, model parameters, and processed inputs.
The adversary observes physical side-channel leakage, such as timing behavior, power consumption, or electromagnetic (EM) emissions, during inference or training to gain information about the neural network.
For example, timing information can reveal activation functions~\cite{batina2019csi}, while power or EM measurements can expose information about processed inputs~\cite{batina2019poster} or architectural properties such as layer types~\cite{yan2023mercury}.

Of particular relevance to our work is correlation power analysis (CPA)~\cite{batina2019csi,puvskavc2025make,lehocky2025side}, in which the correlation between hypothetical leakage values derived from candidate secret parameters and measured side-channel leakage is evaluated.
By identifying the parameter hypothesis that yields the strongest correlation with the observed leakage, an attacker can infer secret model parameters.
Such attacks rely on the dependence of physical leakage on processed data and sensitive parameters, which gives rise to statistically distinguishable leakage behavior.

\subsection{Leakage Assessment}
\label{sec:la}
Leakage assessment provides an implementation-level methodology for determining whether side-channel measurements exhibit statistically detectable data-dependent leakage, without requiring knowledge of a specific attack strategy.
Because exhaustively evaluating an implementation against all possible attack strategies is impractical, leakage assessment provides a useful preliminary security evaluation by statistically analyzing measured side-channel leakage~\cite{standaert2016towards,hou2024cryptography}.

Among the proposed methodologies, Test Vector Leakage Assessment (TVLA)~\cite{gilbert2011testing} has become a widely adopted methodology for evaluating leakage in cryptographic implementations.
TVLA employs statistical hypothesis testing to detect leakage. 
It has also been adopted to evaluate the side-channel security of neural network implementations.
To the best of our knowledge, TVLA remains the only established leakage assessment methodology currently applied to neural network implementations. 
Further details on TVLA and its statistical foundations are provided in Subsections~\ref{sec:welch} and~\ref{sec:SHTandLA}.

Leakage vulnerabilities can also be assessed earlier in the hardware-design flow.
For example, SCAR~\cite{srivastava2024scar} performs pre-silicon power side-channel analysis at the RTL level and uses graph neural networks to detect and localize modules susceptible to side-channel leakage.
Such design-time approaches complement measurement-based methodologies such as TVLA and the statistical leakage-assessment techniques considered in this work.
 
Beyond TVLA, Moradi et al.~\cite{moradi2018leakage} proposed Pearson's $\chi^2$ test for leakage detection, allowing differences between leakage populations to be detected beyond differences in their means.
Applying the \(\chi^2\) test to numerical side-channel measurements requires discretization of the measurement space, making the resulting assessment dependent on choices such as the binning strategy and number of bins.

Apart from statistical tests, Moos et al.~\cite{moos2021dl} proposed Deep Learning Leakage Assessment (DL-LA), in which a neural network classifier is trained to distinguish between two groups of side-channel traces. 
DL-LA can detect complex, multivariate, and misaligned leakage patterns without requiring explicit specification of their statistical order.
However, it requires separate training and validation data sets as well as neural network training, increasing the evaluation effort compared with lightweight statistical leakage-assessment methods.

\subsection{Countermeasures Against Side-Channel Analysis Attacks}
\label{sec:countermeasure}
Several countermeasures have been proposed to mitigate SCA attacks on neural network implementations.
Desynchronization-based techniques introduce timing variability into the computation to reduce the temporal alignment of side-channel measurements.
One common approach is random jitter, in which variable delays are inserted during execution so that the same intermediate computation does not consistently occur at the same time position across traces~\cite{coron2009efficient}.
Such desynchronization techniques have also been investigated for neural network implementations~\cite{breier2023desynchronization}.
Masking approaches~\cite{dubey2022modulonet} randomize intermediate computations to reduce the statistical dependence between measured leakage and sensitive variables.
Such approaches have been demonstrated for neural networks with integer weights to hinder CPA.
Protecting intermediate computations through masking can, however, introduce substantial computational and implementation overhead.
Encoding-based countermeasures provide another software-level approach and have been investigated for improving side-channel resistance in cryptographic implementations~\cite{hou2024another}.
Although encoding-based protection has not yet been extensively studied for mitigating SCA attacks on neural network implementations, it has recently been applied to neural networks as a countermeasure against fault attacks~\cite{velcicky2026deepncode}.
Alternative approaches modify the inference pipeline itself; for example, Parasite~\cite{chabanne2021parasite} applies additional convolutional models to the victim model's input to hinder EM-based parameter extraction.

The implementation evaluated in this work employs shuffling to hinder CPA by randomizing the temporal location of weight-related computations.
Shuffling randomizes the execution order of these computations within a layer, thereby reducing the temporal correspondence between predicted intermediate values and measured leakage~\cite{nozaki2021shuffling,brosch2022counteract}.
Subsequent work further protected the shuffled-index generation mechanism itself against SCA~\cite{puvskavc2025make}.

\section{Statistical Preliminaries for Leakage Assessment}
\label{sec:preliminaries}
This section introduces the statistical notation used throughout the paper (Subsection~\ref{sec:notation}), reviews Welch's $t$-test underlying TVLA (Subsection~\ref{sec:welch}), and describes the two-sample Anderson--Darling test underlying ADLA (Subsection~\ref{sec:AD}).

\subsection{Notation and Preliminaries}
\label{sec:notation}
A statistical hypothesis~\cite{ross2020introduction} is a statement about the probability distribution governing observed data and is evaluated using samples drawn from that distribution.

\textit{Statistical hypothesis testing} is a methodological framework that uses sample data to evaluate such statements.
More precisely, it provides a decision rule for determining whether the observed sample is consistent with a specified hypothesis about the underlying data-generating distribution(s).
Based on the outcome of the test, the hypothesis is either rejected or not rejected.
Failure to reject a hypothesis does not imply that it is true; rather, it indicates that the observed data do not provide sufficient evidence against it.

The hypothesis under investigation is referred to as the \textit{null hypothesis}, denoted by $H_0$.
It is tested against a competing statement called the \textit{alternative hypothesis}, denoted by $H_1$. The performance of the test is characterized by its \textit{significance level}, denoted by $\alpha$, which is defined as an upper bound on the probability of rejecting $H_0$ when $H_0$ is true (Type~I error).
For a given choice of \(\alpha\), a critical region is determined according to the distribution of the test statistic under \(H_0\).

In this paper, we focus on the comparison of two probability distributions. 
Let $X$ and $Y$ denote the corresponding random variables.
Let $\{X_1, X_2, \dots, X_n\}$ and $\{Y_1, Y_2, \dots, Y_n\}$ be independent samples drawn from the distributions of $X$ and $Y$, respectively\footnote{For simplicity, we assume that both samples have the same size $n$.
This assumption is justified in the context of side-channel measurements, where it is typically feasible to collect an equal number of traces under different experimental conditions.}.

A test statistic is computed from the observed samples.
If the value of the test statistic falls within the critical region, the null hypothesis \(H_0\) is rejected; otherwise, it is not rejected.

\subsection{Welch's $t$-test}
\label{sec:welch}
Welch's \(t\)-test~\cite{welch1947generalization} is a parametric test for comparing the means of two independent normal populations without assuming equal variances.

Let $X \sim \mathcal{N}(\mu_x,\sigma_x^2)$ and $Y \sim \mathcal{N}(\mu_y,\sigma_y^2)$ denote two independent random variables corresponding to the distributions under consideration. 
The null and alternative hypotheses are defined as
\[
H_0:\mu_x=\mu_y,
\qquad
H_1:\mu_x\neq\mu_y.
\]
The test statistic is given by
\begin{equation}\label{eq:t}
    t :=
\frac{\overline{X}-\overline{Y}}
{\sqrt{\frac{S_x^2}{n}+\frac{S_y^2}{n}}},
\end{equation}
where $\overline{X}$ and $\overline{Y}$ denote the sample means, and $S_x^2$ and $S_y^2$ denote the unbiased sample variances of the respective samples.

Under the null hypothesis, the statistic $t$ approximately follows a $t$-distribution.
For sufficiently large sample sizes, the distribution of $t$ converges to the standard normal distribution by the central limit theorem.
In this asymptotic regime, the threshold corresponding to a significance level $\alpha$ is $z_{1-\alpha/2}$, defined by
\[
\Phi(z_{1-\alpha/2}) = 1 - \frac{\alpha}{2},
\]
where $\Phi$ denotes the cumulative distribution function of the standard normal distribution.
Equivalently,
\begin{equation}\label{eq:zalpha}
    \frac{\alpha}{2} = 1 - \Phi(z_{1-\alpha/2}).
\end{equation}
The null hypothesis is rejected if $|t| > z_{1-\alpha/2}$.
In this case, at significance level $\alpha$, the observed data provide sufficient statistical evidence to reject $H_0$ in favor of $H_1$, indicating that the population means differ.

\subsection{Two-sample Anderson--Darling Test}
\label{sec:AD}

The two-sample Anderson--Darling test~\cite{pettitt1976two} is a nonparametric procedure for testing whether two independent samples originate from the same (continuous) distribution. In contrast to mean-based tests, it compares the entire distributions via their cumulative distribution functions.

Let $F_x$ and $F_y$ denote the cumulative distribution functions (CDFs) of the random variables $X$ and $Y$, respectively. The null and alternative hypotheses are
\[
H_0: F_x = F_y,
\qquad
H_1: F_x \neq F_y.
\]
Let $\{X_1,\dots,X_n\}$ and $\{Y_1,\dots,Y_n\}$ be two independent samples of equal size $n$. Consider the pooled sample of size $2n$, arranged in increasing order,
\[
Z_{(1)} \le Z_{(2)} \le \dots \le Z_{(2n)}.
\]
The pooled ordering retains the sample label of each observation and therefore captures how observations from the two populations are interleaved across the empirical distribution.
For each $i \in \{1,\dots,2n-1\}$, let $M_i$ denote the number of observations among $\{X_1,\dots,X_n\}$ that are less than or equal to $Z_{(i)}$.

For two equally sized samples, the Anderson--Darling statistic can be written as\footnote{The square in the notation is historical and reflects the fact that the statistic is a quadratic functional of the empirical process.}
\begin{equation}\label{eq:A2}
    A^2 :=
\frac{1}{n^2}
\sum_{i=1}^{2n-1}
\frac{(2n M_i - n i)^2}{i(2n-i)}.
\end{equation}

Under $H_0$ and assuming a continuous common distribution, the statistic $A^2$ converges in distribution as $n \to \infty$. 
For the setting considered here, the limiting random variable can be expressed as~\cite{scholz1987k}
\begin{equation}\label{eq:Ainfty}
    A^2_\infty
:=
\sum_{j=1}^{\infty}
\frac{1}{j(j+1)} W_j,
\end{equation}
where $\{W_j\}_{j\ge1}$ are independent $\chi^2$ random variables with one degree of freedom.

Since the limiting distribution does not admit a closed-form expression, thresholds corresponding to a prescribed significance level $\alpha$ are obtained from tabulated asymptotic percentiles or via numerical approximation. 
Scholz and Stephens~\cite{scholz1987k} approximated these percentiles by matching the first four moments of the limiting distribution and fitting a Pearson curve, following Stephens~\cite{stephens1976asymptotic} and Solomon and Stephens~\cite{solomon1978approximations}. 
We will use the same approach in Subsection~\ref{sec:threshold} to derive the ADLA decision threshold.

\section{Anderson--Darling Leakage Assessment}
\label{sec:ADLA}
This section introduces the ADLA framework for leakage assessment of neural network implementations. Subsection~\ref{sec:SHTandLA} formulates leakage detection as a statistical hypothesis-testing problem and describes how TVLA is applied in this setting. Subsection~\ref{sec:threshold} then introduces ADLA as a distribution-sensitive alternative to mean-based leakage assessment and derives its decision threshold based on the two-sample Anderson--Darling test introduced in Subsection~\ref{sec:AD}.

\subsection{Leakage Detection as a Statistical Hypothesis Test}
\label{sec:SHTandLA} 
As discussed in Section~\ref{sec:scaonnn}, SCAs targeting neural networks can recover secret model parameters by exploiting statistical dependencies between intermediate computations and the corresponding side-channel observations. 
Leakage assessment aims to determine whether statistically detectable data-dependent leakage is present in such computations without requiring a specific parameter-recovery attack.

To formalize leakage detection within the framework of statistical hypothesis testing, we model side-channel measurements obtained under different experimental conditions as realizations of random variables.  

In our evaluation, we consider two input configurations for leakage assessment: fixed-versus-fixed (FvF) and fixed-versus-random (FvR).
In both cases, the measurements are partitioned into two populations according to the applied input condition.
In the FvF setting, the two populations correspond to two distinct fixed input configurations, whereas in the FvR setting, one population corresponds to a fixed input configuration and the other to randomized inputs.

In the considered neural network implementation, we target a specific model weight and vary the input value that is multiplied by this weight during inference, while keeping all remaining input components fixed.
Specifically, we target the first weight of the first hidden neuron, vary the associated input value (i.e. the first input value), and leave all remaining input values unchanged.
In the FvF setting, the first input feature assumes one of two fixed values.
In the FvR configuration, one population uses a fixed value of the first input feature, whereas the other uses values independently drawn from a uniform distribution over the normalized input range \([0,1]\).
This corresponds to a controlled chosen-input setting in which the selected input value can be varied to induce changes in the intermediate computation involving the targeted weight.

Under the null hypothesis of no data-dependent leakage, the distribution of the measured leakage should be invariant with respect to the applied input condition.
Consequently, the leakage distributions obtained from the two measurement populations should be identical, regardless of whether the comparison is performed in the FvF or FvR setting.
A statistically significant difference between these distributions therefore provides evidence that the measured side-channel leakage depends on the selected input condition.

TVLA evaluates this dependence by comparing the means of the two leakage populations using Welch's $t$-test~\cite{hou2024cryptography}.
A statistically significant difference between the means provides evidence of data-dependent leakage.
Conversely, failure to detect a significant mean difference does not imply that the two leakage distributions are identical; differences in other distributional properties may still be present.

\subsection{Derivation of the ADLA Threshold}
\label{sec:threshold}
Motivated by the limitations of mean-based leakage assessment, we propose \emph{Anderson--Darling Leakage Assessment (ADLA)}, which employs the two-sample Anderson--Darling test to compare the leakage distributions more broadly.
In contrast to Welch's $t$-test, the Anderson--Darling test is sensitive to distributional differences beyond changes in the mean, including differences in spread, shape, and tail behavior.

In standard TVLA practice, the detection threshold is set to $\tau_t:=4.5$~\cite{gilbert2011testing,ding2018towards}. That is, after computing the test statistic $t$ (cf.~Eq.~\ref{eq:t}), leakage is declared if
\[
|t| > \tau_t=4.5.
\]
Under the asymptotic normal approximation (cf.~Eq.~\ref{eq:zalpha}), this threshold corresponds to a two-sided significance level of approximately
\[
\alpha \approx 6.8 \times 10^{-6},
\]
providing a highly conservative criterion for leakage detection~\cite{hou2024cryptography}.

To ensure direct comparability with the standard TVLA methodology, we calibrate ADLA to the same nominal significance level \(\alpha\).

As discussed in Subsection~\ref{sec:AD}, the limiting distribution of the two-sample Anderson--Darling statistic does not admit a closed-form expression. 
Consequently, the critical value corresponding to the prescribed significance level \(\alpha\) must be obtained numerically.
We denote the threshold for ADLA by $\tau_A$, defined as the upper $(1-\alpha)$-quantile of the limiting distribution:
\[
\Pr\!\left( A^2_\infty > \tau_A \right) = \alpha\approx6.8\times10^{-6}.
\]
ADLA therefore declares leakage when
\[
A^2>\tau_A.
\]
To approximate $\tau_A$, we employ the Pearson curve fitting method~\cite{scholz1987k}. 
From Eq.~\ref{eq:Ainfty} and the additivity and scaling properties of cumulants for independent random variables~\cite{kolassa2006series}, the $r$th cumulant of $A^2_\infty$ is
\[
\kappa_r = 2^{r-1}(r-1)! \sum_{j=1}^{\infty} \frac{1}{\bigl(j(j+1)\bigr)^r}.
\]
For the first four cumulants, the corresponding series can be evaluated in closed form using partial-fraction decomposition~\cite{hata2007problems}:
\[
\sum_{j=1}^{\infty}\frac{1}{(j(j+1))^r}
=
\begin{cases}
1, & r=1,\\[6pt]
\dfrac{\pi^2}{3}-3, & r=2,\\[8pt]
10-\pi^2, & r=3,\\[8pt]
\dfrac{\pi^4}{45}+\dfrac{10\pi^2}{3}-35, & r=4.
\end{cases}
\]
Substituting these expressions into the cumulant formula yields
\[
\kappa_1 = 1,\quad
\kappa_2 = \frac{2\pi^2}{3}-6,
\]
\[
\kappa_3 = 80-8\pi^2,\quad
\kappa_4 = \frac{16\pi^4}{15}+160\pi^2-1680.
\]
Using the standard relations between cumulants and central moments~\cite{mittelhammer2013mathematical}, the first four moments satisfy
\[
\mu = \kappa_1,\qquad
\mu_2 = \kappa_2,\qquad
\mu_3 = \kappa_3,\qquad
\mu_4 = \kappa_4 + 3\kappa_2^2.
\]
The skewness and kurtosis are therefore given by~\cite{sharma2012business}
\[
\gamma_1 = \frac{\mu_3}{\mu_2^{3/2}},
\qquad
\beta_2 = \frac{\mu_4}{\mu_2^2}.
\]
The mean \(\mu\), variance \(\mu_2\), skewness \(\gamma_1\), and kurtosis \(\beta_2\) determine the corresponding Pearson-system approximation, which we use to approximate the limiting distribution of \(A^2_\infty\).
The Pearson curve fitting is performed using the \texttt{PearsonDS} package~\cite{becker2025pearsonds} in \textsf{R}~\cite{R}. 
For $\alpha = 6.8 \times 10^{-6}$, the resulting ADLA threshold value is
\[
\tau_A \approx 11.18.
\]
Hence, throughout the remainder of this work, ADLA declares leakage when $A^2>11.18$.

\section{Experimental Evaluation}
\label{sec:evaluation}
This section evaluates ADLA on the shuffling-protected neural network implementation and compares its leakage-detection behavior with TVLA.
Section~\ref{sec:experimental_setup} describes the experimental setup, while Section~\ref{sec:leakage_detection} presents the leakage-assessment results.

\subsection{Experimental Setup}
\label{sec:experimental_setup}
The experiments are conducted on an embedded implementation of an MLP trained for handwritten-digit classification on the MNIST dataset~\cite{deng2012mnist}.
The network takes \(784\) input features, corresponding to a flattened \(28\times28\) grayscale image, and contains two fully connected hidden layers with \(300\) and \(100\) neurons, respectively, followed by a \(10\)-neuron output layer.
ReLU activation functions are employed after both hidden layers, while the output layer uses a softmax activation.
The input pixels are normalized to the range \([0,1]\).
The network is trained using the Adam optimizer with a learning rate of \(10^{-3}\), a batch size of \(128\), and \(20\) training epochs.
The trained network achieves a test accuracy of \(98.03\%\) on the MNIST test set.

For the side-channel experiments, the target firmware implements the computation of the first neuron of the first hidden layer.
The neuron therefore computes the weighted sum of its \(784\) input features using the corresponding trained weights, adds the bias, and subsequently applies the ReLU activation.
The neuron computation is performed using 32-bit IEEE-754 single-precision floating-point arithmetic.
To evaluate leakage detection in the presence of execution-order randomization, we employ the shuffling-protected implementation of~\cite{puvskavc2025make}, in which a fresh permutation determines the execution order of the \(784\) multiply-and-accumulate computations for each inference.
The arithmetic operations and model parameters are otherwise unchanged by the protection mechanism.
The firmware is compiled with optimization level \texttt{-O2} and communicates with the host through SimpleSerial~2.1.

Power traces are acquired using a ChipWhisperer-Husky~\cite{chipwhisperer_husky_manual} connected to a CW313 target board equipped with an Atmel SAM4S (ARM Cortex-M4) microcontroller.
The target clock frequency is configured to approximately \(7.364\,\mathrm{MHz}\).
The Husky ADC operates at approximately \(29.455\,\mathrm{MHz}\) with a clock multiplier of \(4\) and an acquisition gain of approximately \(25.09\,\mathrm{dB}\).
A decimation factor of \(2\) is applied during acquisition, yielding an effective stored sampling rate of approximately \(14.727\,\mathrm{MS/s}\).
Each captured trace contains \(120{,}000\) samples.

The measurement trigger covers the weighted accumulation loop and the subsequent ReLU operation.
Consequently, each captured trace contains the full execution interval of the operation under investigation.
Both fixed-versus-fixed (FvF) and fixed-versus-random (FvR) input configurations are considered.
In the FvF setting, the first input feature is assigned two distinct fixed values, one for each measurement group.
In the FvR setting, one group uses a fixed value of the first input feature, while the other group uses values independently resampled for each trace from a uniform distribution over \([0,1]\).
The first model weight is selected as the target, and only the input feature associated with this weight is varied, while the remaining \(783\) input features are taken from a fixed normalized MNIST test sample and held constant throughout the experiment.

Each experimental setting therefore consists of two measurement groups that differ only in the value assigned to the input feature associated with the targeted first weight.
For each experimental setting, \(25{,}000\) traces are collected for each of the two comparison groups, resulting in \(50{,}000\) traces in total.
Traces from the two comparison groups are acquired in an interleaved manner, with measurements from the two groups alternated during acquisition rather than collected in separate sequential batches.
This acquisition strategy reduces the influence of slow temporal drift in the measurement setup on the comparison between the two groups.

\subsection{TVLA and ADLA Leakage Detection}
\label{sec:leakage_detection}

Although each acquired trace contains \(120{,}000\) samples, the leakage analysis is restricted to samples \(0\) through \(108{,}299\), corresponding to the first \(108{,}300\) samples.
This interval is determined from the trigger duration reported by the ChipWhisperer acquisition hardware through \texttt{scope.adc.trig\_count} and corresponds to the execution interval of the triggered first-neuron computation.
Samples following the end of the trigger are therefore excluded from the statistical analysis.

At each time sample, we compute the absolute Welch \(t\)-statistic \(|t|\) for TVLA and the two-sample Anderson--Darling statistic \(A^2\) for ADLA, as defined in Eqs.~\eqref{eq:t} and~\eqref{eq:A2}, respectively.
For ADLA, the statistic is implemented directly according to Eq.~\eqref{eq:A2}, rather than using a library-specific implementation or standardized variant of the Anderson--Darling statistic.
This ensures consistency between the statistic used in the experiments and the statistic for which the decision threshold \(\tau_A\) was derived in Section~\ref{sec:threshold}.

To evaluate how the leakage-detection behavior evolves with the number of measurements, both statistics are recomputed for \(100\) progressively increasing trace counts,
\[
n \in \{250,500,750,\ldots,25{,}000\},
\]
where \(n\) denotes the number of traces used from each of the two comparison groups.
For each value of \(n\), the first \(n\) traces from each group are used.
The final point, \(n=25{,}000\), corresponds to the complete assessment using all \(25{,}000\) traces from each group.

For each \(n\), the TVLA and ADLA statistics are evaluated at every time sample in the analyzed trace interval.
To summarize the strongest detected leakage at a given trace count, we consider the maximum statistic over all analyzed time samples,
\[
t_{\max}(n)
=
\max_s |t_s(n)|,
\qquad
A^2_{\max}(n)
=
\max_s A^2_s(n),
\]
where \(s\in\{0,1,\dots,108299\}\) denotes the time-sample index.
For visualization, these maxima are normalized by their respective decision thresholds,
\[
\frac{t_{\max}(n)}{\tau_t}
\qquad\text{and}\qquad
\frac{A^2_{\max}(n)}{\tau_A},
\]
where \(\tau_t=4.5\) and \(\tau_A=11.18\).
A normalized value greater than \(1\) indicates that at least one time sample exceeds the corresponding pointwise decision threshold at the nominal significance level \(\alpha=6.8\times10^{-6}\).
In the leakage-assessment setting considered here, such a threshold exceedance is interpreted as statistical evidence of data-dependent leakage at the corresponding time sample.

\begin{figure}[htb]
    \centering

    \begin{tabular}{cc}
        \includegraphics[width=0.48\linewidth]{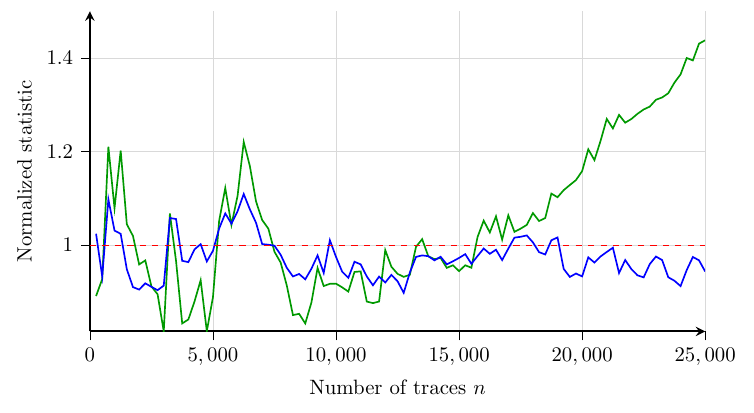} &
        \includegraphics[width=0.48\linewidth]{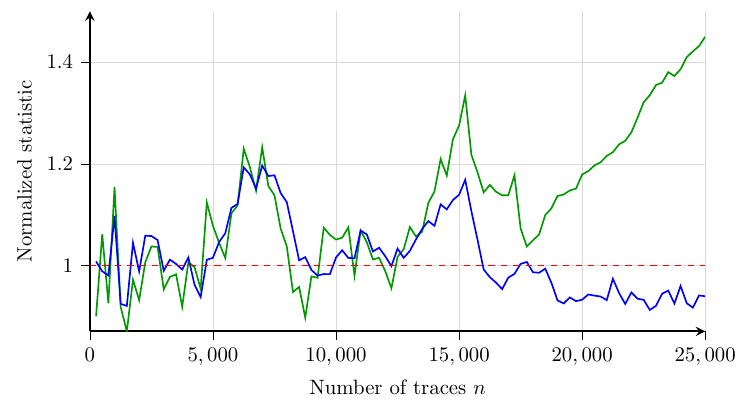} \\

        \includegraphics[width=0.48\linewidth]{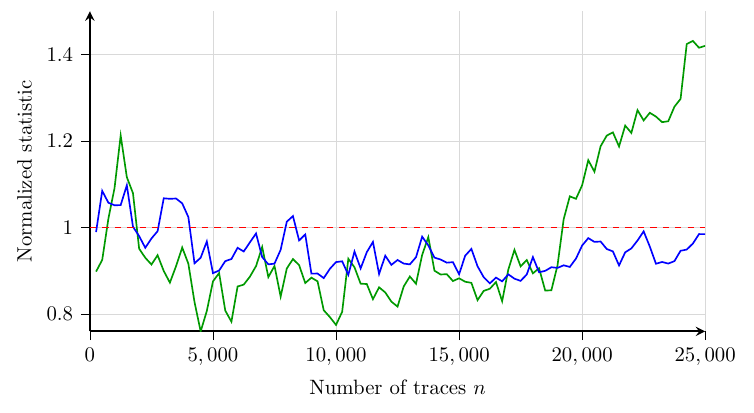} &
        \includegraphics[width=0.48\linewidth]{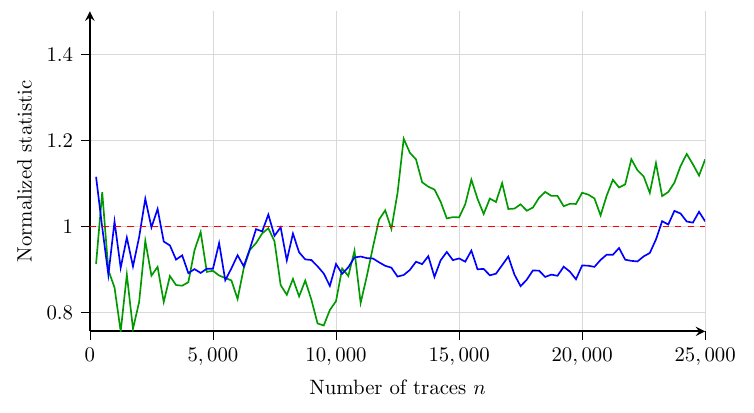}
    \end{tabular}

    \vspace{0.5em}

    \begin{tikzpicture}[baseline=-0.6ex]
        \draw[green!60!black, thick] (0,0) -- (0.7,0);
        \node[anchor=west] at (0.8,0)
            {$\frac{A^2_{\max}}{\tau_A}$ (ADLA)};

        \draw[blue, thick] (3.8,0) -- (4.5,0);
        \node[anchor=west] at (4.6,0)
            {$\frac{t_{\max}}{\tau_t}$ (TVLA)};

        \draw[red, dashed, thin] (7.5,0) -- (8.2,0);
        \node[anchor=west] at (8.3,0)
            {threshold \(1\)};
    \end{tikzpicture}

    \caption{Threshold-normalized maximum ADLA and TVLA statistics as a function of the number of traces per group.
    The top row shows two fixed-versus-random (FvR) settings, while the bottom row shows two fixed-versus-fixed (FvF) settings.
    For each trace count \(n\), the first \(n\) traces from each comparison group are used and the maximum statistic over the analyzed trace interval is reported.
    The horizontal dashed line denotes the normalized detection threshold of \(1\).}
    \label{fig:assessment}
\end{figure}

Figure~\ref{fig:assessment} summarizes the normalized ADLA and TVLA results for the two FvR and two FvF settings.
At lower trace counts, both statistics exhibit fluctuations around the detection threshold.
As the number of traces increases, however, ADLA develops more pronounced threshold exceedances.
From approximately \(20{,}000\) traces onward, the ADLA statistic shows persistent threshold exceedances across the evaluated settings, whereas the TVLA statistic generally remains close to the decision threshold even at \(25{,}000\) traces.
This indicates that, for the considered implementation, ADLA provides a clearer separation from its decision threshold than TVLA.

\begin{figure}[htb]
    \centering

    \begin{tabular}{cc}
        \includegraphics[width=0.48\linewidth]{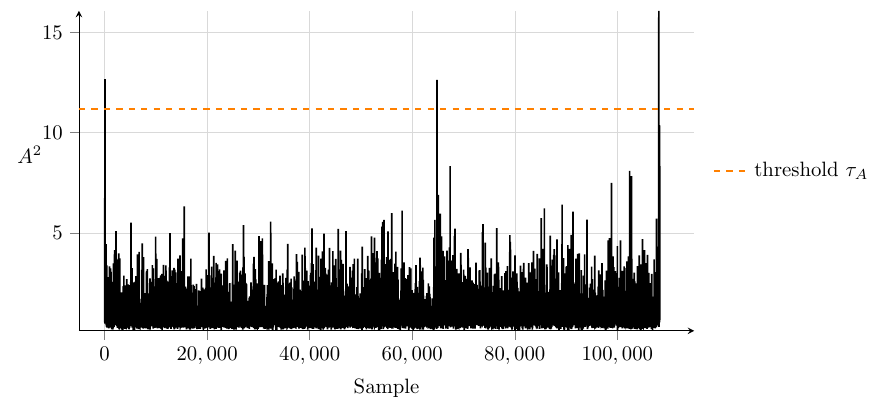} &
        \includegraphics[width=0.48\linewidth]{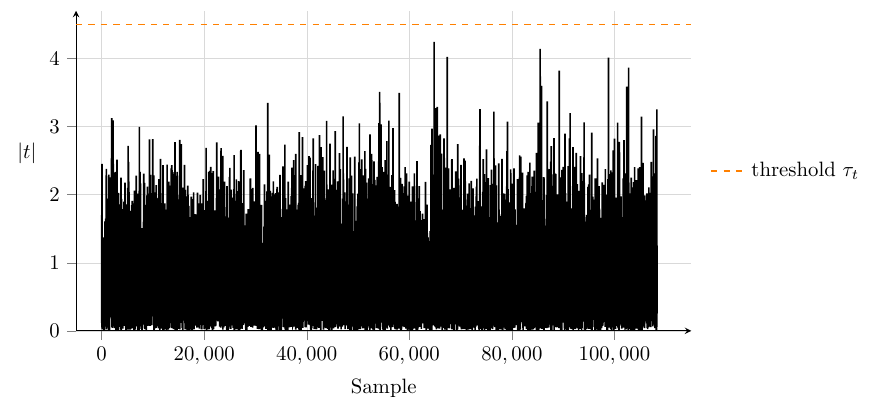} \\
        ADLA (\(A^2\)) & TVLA (\(|t|\))
    \end{tabular}

    \caption{Pointwise ADLA and TVLA statistics across the analyzed trace interval for a representative FvR setting using \(25{,}000\) traces per group.
    The horizontal lines denote the corresponding decision thresholds, \(\tau_A=11.18\) for ADLA and \(\tau_t=4.5\) for TVLA.}
    \label{fig:tvla_ad}
\end{figure}

To examine the pointwise behavior underlying the convergence curves, Figure~\ref{fig:tvla_ad} shows the ADLA and TVLA statistics across the complete analyzed trace interval for a representative FvR setting using all \(25{,}000\) traces from each group.
In this setting, the TVLA statistic \(|t|\) remains below its decision threshold throughout the analyzed interval.
In contrast, the ADLA statistic \(A^2\) exceeds its threshold at three time samples, providing pointwise statistical evidence of leakage at locations that are not identified by TVLA.

Overall, the experimental results indicate that ADLA is more sensitive to the leakage present in the evaluated settings, identifying statistically significant distributional differences that are not detected by TVLA.
The difference is reflected not only in whether the respective decision thresholds are exceeded, but also in the margin relative to these thresholds.
The normalized TVLA statistic generally remains close to \(1\), whereas the normalized ADLA statistic exhibits larger threshold exceedances as the number of traces increases.
These observations suggest that the differences between the leakage distributions are not fully characterized by changes in their means and may instead involve other distributional properties to which ADLA is sensitive but which are not captured by the mean-based TVLA statistic.

\section{CPA Evaluation}
\label{sec:cpa}
To evaluate whether the leakage identified by ADLA can be exploited in a practical side-channel attack, we perform a CPA targeting the first weight of the first hidden-layer neuron.
The attack follows a methodology similar to that of~\cite{puvskavc2025make} and uses the points of interest identified by ADLA and TVLA, respectively.
We focus on recovery of the exponent byte of the target weight represented in IEEE-754 single-precision format.

\begin{figure}[htb]
    \centering

    \begin{tabular}{cc}
        \includegraphics[width=0.48\linewidth]{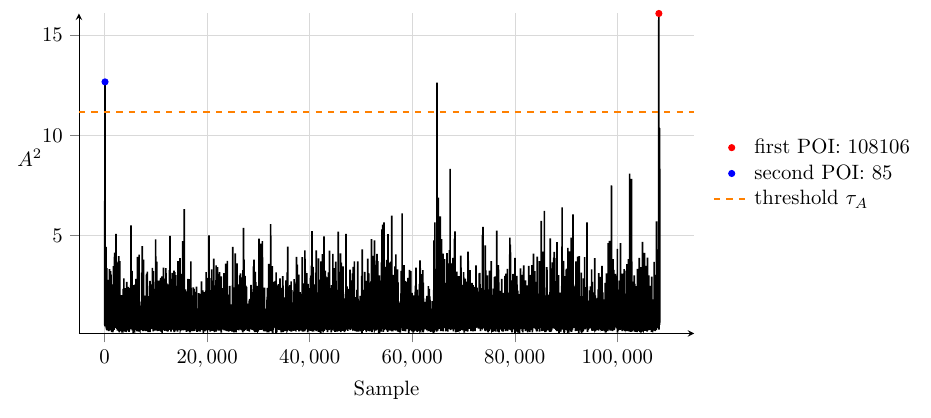} &
        \includegraphics[width=0.48\linewidth]{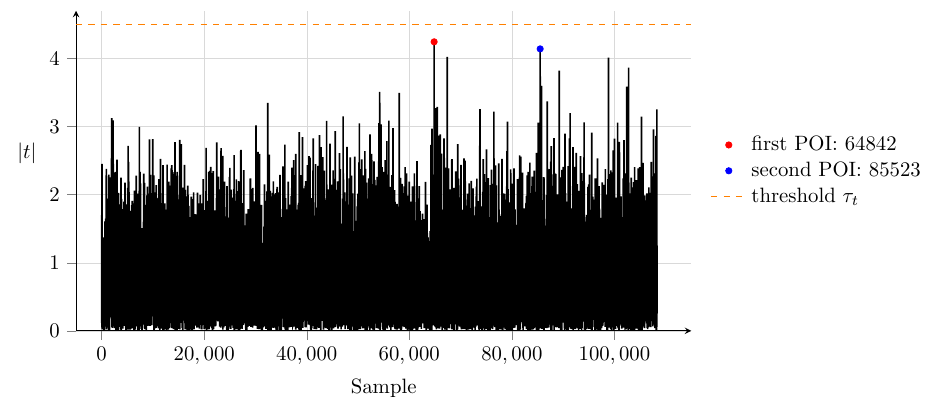} \\
        ADLA & TVLA
    \end{tabular}

    \caption{Points of interest selected from the ADLA and TVLA leakage-assessment statistics.
    The two most prominent regional peaks are selected for each method and subsequently used in the CPA evaluation.}
    \label{fig:POIs}
\end{figure}

\noindent
\textbf{POI Selection.}
The POIs are selected from the full-trace ADLA and TVLA statistics shown in Figure~\ref{fig:tvla_ad}.
For each method, regional peaks are detected in the corresponding statistic, and the two most prominent peaks at a minimum separation of \(5{,}000\) samples are selected.
This procedure isolates the strongest localized responses while avoiding the selection of multiple closely spaced samples belonging to the same peak region.
The resulting POIs therefore correspond to the two dominant regional peaks in the ADLA and TVLA curves, respectively.
For ADLA, the selected POIs are samples \(108106\) and \(85\), whereas TVLA identifies samples \(64842\) and \(85523\).
The selected POIs are highlighted in Figure~\ref{fig:POIs}.

\noindent
\textbf{Attack Setting.}
The CPA evaluation uses a pool of \(50{,}000\) traces.
For each trace the first input feature is independently drawn from a uniform distribution over \([0,1]\), while the remaining \(783\) input features are held constant as described in Section~\ref{sec:experimental_setup}.

We assume that the attacker knows that the target weight lies within the interval \([-0.5,0.5]\).
Candidate weights are generated with a step size of \(0.0001\), with the value zero excluded, resulting in \(10{,}000\) candidate weight values.

We adopt a second-order CPA approach in which the leakage values at the two selected POIs are combined by multiplication before computing the correlation.
For trace \(i\), let \(\ell_i[s]\) denote the measured leakage at sample \(s\).
The leakage values associated with the ADLA-selected POIs are combined as
\[
L_i^{\mathrm{ADLA}}
=
\ell_i[108106]\cdot\ell_i[85],
\]
while the leakage values associated with the TVLA-selected POIs are combined as
\[
L_i^{\mathrm{TVLA}}
=
\ell_i[64842]\cdot\ell_i[85523].
\]
This results in two combined leakage vectors, \(L^{\mathrm{ADLA}}\) and \(L^{\mathrm{TVLA}}\), each containing \(50{,}000\) entries.
For each candidate weight \(w\) and known input value \(x_i\), the hypothetical leakage is modeled as the Hamming weight of their product using the \(32\)-bit IEEE-754 single-precision representation,
\[
H_i(w)
=
\operatorname{HW}\!\left(wx_i\right).
\]
For each candidate weight, the absolute Pearson correlation is computed between the hypothetical leakage vector \(H(w)\) and the corresponding second-order leakage vector, \(L^{\mathrm{ADLA}}\) or \(L^{\mathrm{TVLA}}\).

In this evaluation, the target is the exponent byte of the candidate weight, corresponding to bits \(30\) through \(23\) of its IEEE-754 representation.
Following a methodology similar to that in~\cite{puvskavc2025make}, candidate weights sharing the same exponent byte are associated with the same exponent hypothesis.
For each possible exponent value \(e\), its CPA score, \(S(e)\), is defined as the maximum absolute correlation among all candidate weights whose exponent byte equals \(e\),
\[
S(e)
=
\max_{w:\,\text{exponent byte of }(w)=e}
\left|
\rho\!\left(H(w),L\right)
\right|,
\]
where \(L\) is either \(L^{\mathrm{ADLA}}\) or \(L^{\mathrm{TVLA}}\).
The exponent byte of the target weight is \(123\) (\(\mathtt{0x7B}\)).

To evaluate the attack as a function of the number of available traces, CPA is performed for progressively increasing trace counts between \(1{,}000\) and \(50{,}000\).
To reduce the dependence of the result on a particular ordering of the measurements, the complete \(50{,}000\)-trace pool is independently permuted \(10\) times.
For each permutation and each trace count \(n\), CPA is performed using the first \(n\) traces of that permutation.
The rank of the correct exponent value is determined from the CPA scores, with rank \(1\) corresponding to the highest-scoring exponent hypothesis.
The reported rank at each trace count is the average rank of the correct exponent byte over the \(10\) permutations.

\begin{figure}[htb]
    \centering
    \includegraphics[width=0.9\linewidth]{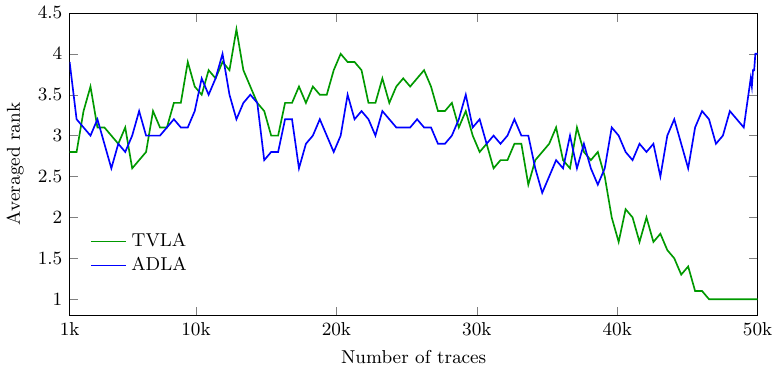}
    \caption{Average rank of the correct exponent byte as a function of the number of traces for CPA using the two POIs identified by ADLA and TVLA, respectively.
    The reported rank is averaged over \(10\) independent permutations of the \(50{,}000\)-trace pool.}
    \label{fig:rank}
\end{figure}

\noindent
\textbf{Attack Results.}
Figure~\ref{fig:rank} shows the averaged rank of the correct exponent byte as a function of the number of traces for the POIs identified by ADLA and TVLA.
The ADLA-selected POIs lead to a clear decrease in the rank of the correct exponent hypothesis as the number of traces increases and ultimately allow the correct exponent byte to reach rank \(1\).
In contrast, the rank obtained using the TVLA-selected POIs remains higher over the evaluated trace range.
These results demonstrate that the additional leakage identified by ADLA is not only statistically detectable but can also provide useful information for recovering the exponent byte of a secret model weight in the presence of the shuffling countermeasure.

\section{Discussion}
\label{sec:discussion}

\noindent
\textbf{Comparison with the \(\chi^2\) Test.}
Pearson's \(\chi^2\) test of homogeneity can also be used to assess whether the leakage observations associated with two input conditions follow the same discretized distribution~\cite{moradi2018leakage}.
At each time sample, the numerical leakage values from the two populations are partitioned into \(K\) bins, producing a \(2\times K\) contingency table of observed bin counts.
Under the null hypothesis, the two input conditions induce the same categorical distribution over these bins, whereas a sufficiently large \(\chi^2\) statistic provides evidence that the corresponding discretized leakage distributions differ.
Unlike TVLA, which is primarily sensitive to differences in the mean, the \(\chi^2\) test can capture more general distributional differences between the two populations.
However, because the leakage observations are numerical rather than inherently categorical, the test requires an explicit discretization of the sample values, and its outcome can therefore depend on both the number of bins and the method used to define their boundaries.

In our evaluation, we consider both equal-width and quantile binning.
For quantile binning, the bin boundaries are determined from the pooled leakage observations of the two groups, while equal-width binning partitions the pooled value range into intervals of equal width.
To avoid unreliable contingency-table cells with small expected counts, adjacent bins are merged until every retained bin has an expected count of at least \(5\).
Consequently, the effective number of bins, denoted by \(K_{\mathrm{eff}}\), may be smaller than the initially selected value \(K\).
For the nominal significance level \(\alpha=6.8\times10^{-6}\), the decision threshold is therefore determined from the chi-square distribution with \(K_{\mathrm{eff}}-1\) degrees of freedom.
Hence, unlike TVLA and ADLA, the \(\chi^2\)-based assessment requires both a discretization choice and a threshold that depends on the resulting effective number of bins.

We apply the \(\chi^2\) test to the same four experimental settings used for the TVLA and ADLA evaluation in Figure~\ref{fig:assessment}.
For each setting, we consider equal-width and quantile binning with \(K \in \{5,10,20,30,50\}\) over the same trace-count sweep from \(250\) to \(25{,}000\) traces per group.
For each trace count and binning configuration, we record the maximum threshold-normalized \(\chi^2\) statistic over the analyzed trace interval.

Table~\ref{tab:chi2_bins} summarizes the dependence of the \(\chi^2\) assessment on the discretization choice.
For each setting and binning strategy, it reports the value of \(K\) that most frequently produces the largest normalized statistic over the trace-count sweep.
The preferred bin count varies across settings and binning strategies, showing that no single value of \(K\) is consistently strongest.
Together with the dependence of the decision threshold on the effective number of bins, this introduces an additional parameter-selection burden compared with TVLA and ADLA.

\begin{table}[htb]
\centering
\caption{Sensitivity of the \(\chi^2\) assessment to the selected bin count.
For each setting and binning strategy, the table reports the value of \(K\) that most frequently produces the largest threshold-normalized \(\chi^2\) statistic over the trace-count sweep.}
\label{tab:chi2_bins}
\begin{tabular}{lccc}
\hline
Setting & Binning & Most frequent \(K\) & Frequency \\
\hline
FvR 1 & Equal-width & 30 & 65\% \\
FvR 2 & Equal-width & 50 & 48\% \\
FvF 1 & Equal-width & 20 & 68\% \\
FvF 2 & Equal-width & 20 & 75\% \\
\hline
FvR 1 & Quantile & 10 & 66\% \\
FvR 2 & Quantile & 10 & 57\% \\
FvF 1 & Quantile & 50 & 65\% \\
FvF 2 & Quantile & 20 & 41\% \\
\hline
\end{tabular}
\end{table}

\noindent
\textbf{Mean-Based and Distribution-Based Leakage Assessment.}
The comparison between TVLA and ADLA highlights an important methodological difference between mean-based and distribution-based leakage assessment.
Welch's \(t\)-test summarizes the distinction between two leakage populations through their means, whereas the Anderson--Darling statistic is sensitive to discrepancies across the empirical distributions.
As a result, the two methods need not react equally to the same physical leakage source.
In particular, a leakage mechanism that changes the spread, shape, or tail behavior of the measured distribution without producing a comparably strong mean shift may result in a pronounced ADLA response while remaining close to the TVLA threshold.
This distinction is especially relevant in protected implementations, where countermeasures such as shuffling can alter the temporal and statistical structure of the observed leakage.

\noindent
\textbf{Practical Considerations.}
From a deployment perspective, ADLA retains a workflow similar to TVLA: both operate directly on two groups of traces and use a fixed decision threshold once the target significance level has been selected.
In contrast, the \(\chi^2\) approach requires an additional discretization step, including the choice of binning strategy and initial bin count, and may further require bin merging to satisfy expected-count requirements.
ADLA therefore provides a distribution-sensitive alternative without introducing an additional discretization parameter.
At the same time, TVLA remains useful as a simple and computationally efficient indicator of mean-dependent leakage, while ADLA can provide complementary information when broader distributional differences are present.

\section{Conclusion}
\label{sec:conclusion}
This work introduced Anderson--Darling Leakage Assessment (ADLA), a leakage assessment methodology based on the two-sample Anderson--Darling test.
To enable direct comparison with TVLA, we derived an ADLA threshold corresponding to the nominal significance level associated with the widely adopted TVLA threshold of \(4.5\).

Experiments on a shuffling-protected embedded MLP showed that ADLA can identify leakage that is not detected as clearly by TVLA in the considered FvF and FvR settings.
The CPA evaluation further showed that POIs selected from the ADLA statistic provide useful information for recovering the exponent byte of the targeted model weight.
Overall, ADLA complements mean-based TVLA by providing sensitivity to broader distributional differences without requiring the discretization choices of histogram-based methods such as the \(\chi^2\) test.

\section*{Acknowledgments}
This work was supported by the Open Research Fund of The State Key Laboratory of Blockchain and Data Security (Grant number: A2566), Zhejiang University.

Generative AI tools were used to assist with language polishing and editorial refinement of the manuscript.

\bibliographystyle{plainnat}
\bibliography{bib}

@article{welch1947generalization,
  title={The generalization of ‘STUDENT'S’problem when several different population varlances are involved},
  author={Welch, Bernard L},
  journal={Biometrika},
  volume={34},
  number={1-2},
  pages={28--35},
  year={1947},
  publisher={Oxford University Press}
}

@inproceedings{standaert2016towards,
  title={Towards Fair and Efficient Evaluations of Leaking Cryptographic Devices: Overview of the ERC Project CRASH, Part I (Invited Talk)},
  author={Standaert, Fran{\c{c}}ois-Xavier},
  booktitle={International Conference on Security, Privacy, and Applied Cryptography Engineering},
  pages={353--362},
  year={2016},
  organization={Springer}
}

@article{hou2021physical,
  title={Physical security of deep learning on edge devices: Comprehensive evaluation of fault injection attack vectors},
  author={Hou, Xiaolu and Breier, Jakub and Jap, Dirmanto and Ma, Lei and Bhasin, Shivam and Liu, Yang},
  journal={Microelectronics Reliability},
  volume={120},
  pages={114116},
  year={2021},
  publisher={Elsevier}
}

@inproceedings{chabanne2021parasite,
  title={Parasite: Mitigating physical side-channel attacks against neural networks},
  author={Chabanne, Herv{\'e} and Danger, Jean-Luc and Guiga, Linda and K{\"u}hne, Ulrich},
  booktitle={International Conference on Security, Privacy, and Applied Cryptography Engineering (SPACE)},
  pages={148--167},
  year={2021},
  organization={Springer}
}

@book{ross2020introduction,
  title={Introduction to probability and statistics for engineers and scientists},
  author={Ross, Sheldon M},
  year={2020},
  publisher={Academic press}
}

@book{mittelhammer2013mathematical,
  title={Mathematical statistics for economics and business},
  author={Mittelhammer, Ron C and Mittelhammer, Ron C},
  year={2013},
  publisher={Springer}
}

@inproceedings{coron2009efficient,
  title={An efficient method for random delay generation in embedded software},
  author={Coron, Jean-S{\'e}bastien and Kizhvatov, Ilya},
  booktitle={Cryptographic Hardware and Embedded Systems-CHES 2009: 11th International Workshop Lausanne, Switzerland, September 6-9, 2009 Proceedings},
  pages={156--170},
  year={2009},
  organization={Springer}
}

@article{dubey2022modulonet,
  title={Modulonet: Neural networks meet modular arithmetic for efficient hardware masking},
  author={Dubey, Anuj and Ahmad, Afzal and Pasha, Muhammad Adeel and Cammarota, Rosario and Aysu, Aydin},
  journal={IACR Transactions on Cryptographic Hardware and Embedded Systems},
  pages={506--556},
  year={2022}
}

@inproceedings{brosch2022counteract,
  title={Counteract side-channel analysis of neural networks by shuffling},
  author={Brosch, Manuel and Probst, Matthias and Sigl, Georg},
  booktitle={2022 Design, Automation \& Test in Europe Conference \& Exhibition (DATE)},
  pages={1305--1310},
  year={2022},
  organization={IEEE}
}

@inproceedings{lehocky2025side,
  title={Side-Channel Analysis of OpenVINO-Based Neural Network Models},
  author={Lehock{\`y}, Zdenko and Breier, Jakub and Jap, Dirmanto and Bhasin, Shivam and Hou, Xiaolu},
  booktitle={International Conference on Availability, Reliability and Security},
  pages={307--324},
  year={2025},
  organization={Springer}
}

@inproceedings{nozaki2021shuffling,
  title={Shuffling countermeasure against power side-channel attack for MLP with software implementation},
  author={Nozaki, Yusuke and Yoshikawa, Masaya},
  booktitle={2021 IEEE 4th International Conference on Electronics and Communication Engineering (ICECE)},
  pages={39--42},
  year={2021},
  organization={IEEE}
}

@inproceedings{breier2023desynchronization,
  title={A desynchronization-based countermeasure against side-channel analysis of neural networks},
  author={Breier, Jakub and Jap, Dirmanto and Hou, Xiaolu and Bhasin, Shivam},
  booktitle={International Symposium on Cyber Security, Cryptology, and Machine Learning},
  pages={296--306},
  year={2023},
  organization={Springer}
}

@article{jayasena2023tvla,
  title={TVLA*: Test vector leakage assessment on hardware implementations of asymmetric cryptography algorithms},
  author={Jayasena, Aruna and Andrews, Emma and Mishra, Prabhat},
  journal={IEEE Transactions on Very Large Scale Integration (VLSI) Systems},
  volume={31},
  number={9},
  pages={1269--1279},
  year={2023},
  publisher={IEEE}
}

@article{puvskavc2025make,
  title={Make Shuffling Great Again: A Side-Channel-Resistant Fisher--Yates Algorithm for Protecting Neural Networks},
  author={Pu{\v{s}}k{\'a}{\v{c}}, Leonard and Benovi{\v{c}}, Marek and Breier, Jakub and Hou, Xiaolu},
  journal={IEEE Transactions on Very Large Scale Integration (VLSI) Systems},
  year={2025},
  publisher={IEEE}
}

@article{yan2023mercury,
  title={Mercury: An automated remote side-channel attack to nvidia deep learning accelerator},
  author={Yan, Xiaobei and Lou, Xiaoxuan and Xu, Guowen and Qiu, Han and Guo, Shangwei and Chang, Chip Hong and Zhang, Tianwei},
  journal={arXiv preprint arXiv:2308.01193},
  year={2023}
}

@manual{chipwhisperer_husky_manual,
  title        = {ChipWhisperer-Husky and HuskyPlus User Manual},
  author       = {{NewAE Technology Inc.}},
  organization = {NewAE Technology Inc.},
  year         = {2025},
  url          = {https://media.newae.com/manuals/cwhusky_cwhuskyplus_manual_may2025.pdf}
}

@inproceedings{batina2019poster,
  title={Poster: Recovering the input of neural networks via single shot side-channel attacks},
  author={Batina, Lejla and Bhasin, Shivam and Jap, Dirmanto and Picek, Stjepan},
  booktitle={Proceedings of the 2019 ACM SIGSAC Conference on Computer and Communications Security},
  pages={2657--2659},
  year={2019}
}

@misc{isoiec17825,
  author       = {{International Organization for Standardization (ISO)} and
                  {International Electrotechnical Commission (IEC)}},
  title        = {{ISO/IEC 17825:2016}: Information technology -- Security techniques -- Testing methods for the mitigation of non-invasive attack classes against cryptographic modules},
  howpublished = {International Standard},
  year         = {2016}
}

@inproceedings{batina2019csi,
  title={$\{$CSI$\}$$\{$NN$\}$: Reverse engineering of neural network architectures through electromagnetic side channel},
  author={Batina, Lejla and Bhasin, Shivam and Jap, Dirmanto and Picek, Stjepan},
  booktitle={28th USENIX Security Symposium (USENIX Security 19)},
  pages={515--532},
  year={2019}
}

@book{goodfellow2016deep,
  title={Deep learning},
  author={Goodfellow, Ian and Bengio, Yoshua and Courville, Aaron and Bengio, Yoshua},
  volume={1},
  year={2016},
  publisher={MIT press Cambridge}
}

@Manual{R,
    title = {PearsonDS: Pearson Distribution System},
    author = {Martin Becker and Stefan Klößner},
    year = {2025},
    note = {R package version 1.3.2},
    url = {https://CRAN.R-project.org/package=PearsonDS},
    doi = {10.32614/CRAN.package.PearsonDS},
}

@book{sharma2012business,
  title={Business statistics},
  author={Sharma, Japuji K},
  year={2012},
  publisher={Pearson Education India}
}

@article{pettitt1976two,
  title={A Two-Sample Anderson--Darling Rank Statistic},
  author={Pettitt, AN},
  journal={Biometrika},
  pages={161--168},
  year={1976},
  publisher={JSTOR}
}

@book{hata2007problems,
  title={Problems and Solutions in Real Analysis},
  author={Hata, M.},
  isbn={9789812776013},
  lccn={2008295629},
  series={Series on number theory and its applications},
  url={https://books.google.sk/books?id=vSxkRgQe0AcC},
  year={2007},
  publisher={World Scientific}
}

@book{kolassa2006series,
  title={Series approximation methods in statistics},
  author={Kolassa, John E},
  year={2006},
  publisher={Springer}
}

@article{deng2012mnist,
  title={The mnist database of handwritten digit images for machine learning research [best of the web]},
  author={Deng, Li},
  journal={IEEE signal processing magazine},
  volume={29},
  number={6},
  pages={141--142},
  year={2012},
  publisher={IEEE}
}

@ARTICLE{velcicky2026deepncode,
  author={Velčický, Patrik and Breier, Jakub and Kovačević, Mladen and Hou, Xiaolu},
  journal={IEEE Transactions on Emerging Topics in Computing}, 
  title={DeepNcode: Encoding-Based Protection Against Bit-Flip Attacks on Neural Networks}, 
  year={2026},
  volume={14},
  number={3},
  pages={1181-1196},
  doi={10.1109/TETC.2026.3704626}
  }

@article{srivastava2024scar,
  title={Scar: Power side-channel analysis at rtl level},
  author={Srivastava, Amisha and Das, Sanjay and Choudhury, Navnil and Psiakis, Rafail and Silva, Pedro Henrique and Pal, Debjit and Basu, Kanad},
  journal={IEEE Transactions on Very Large Scale Integration (VLSI) Systems},
  volume={32},
  number={6},
  pages={1110--1123},
  year={2024},
  publisher={IEEE}
}

@article{hou2024another,
  title={Another look at side-channel-resistant encoding schemes},
  author={Hou, Xiaolu and Breier, Jakub and Kova{\v{c}}evi{\'c}, Mladen},
  journal={IEEE Transactions on Very Large Scale Integration (VLSI) Systems},
  volume={32},
  number={8},
  pages={1559--1563},
  year={2024},
  publisher={IEEE}
}

@inproceedings{ding2018towards,
  title={Towards sound and optimal leakage detection procedure},
  author={Ding, A Adam and Zhang, Liwei and Durvaux, Fran{\c{c}}ois and Standaert, Fran{\c{c}}ois-Xavier and Fei, Yunsi},
  booktitle={Smart Card Research and Advanced Applications: 16th International Conference, CARDIS 2017, Lugano, Switzerland, November 13--15, 2017, Revised Selected Papers},
  pages={105--122},
  year={2018},
  organization={Springer}
}

@article{moos2021dl,
  title={DL-LA: Deep learning leakage assessment: A modern roadmap for SCA evaluations},
  author={Moos, Thorben and Wegener, Felix and Moradi, Amir},
  journal={IACR Transactions on Cryptographic Hardware and Embedded Systems},
  pages={552--598},
  year={2021}
}

@article{moradi2018leakage,
  title={Leakage detection with the $\chi^2$-test},
  author={Moradi, Amir and Richter, Bastian and Schneider, Tobias and Standaert, Fran{\c{c}}ois-Xavier},
  volume={2018},  
  number={1}, 
  journal={IACR Transactions on Cryptographic Hardware and Embedded Systems}, 
  year={2018}, 
  pages={209--237} 
}

@inproceedings{gilbert2011testing,
  title={A testing methodology for side-channel resistance validation},
  author={Gilbert Goodwill, Benjamin Jun and Jaffe, Josh and Rohatgi, Pankaj and others},
  booktitle={NIST non-invasive attack testing workshop},
  volume={7},
  pages={115--136},
  year={2011}
}

@book{hou2024cryptography,
  title={Cryptography and Embedded Systems Security},
  author={Hou, Xiaolu and Breier, Jakub},
  year={2024},
  publisher={Springer}
}

@article{stephens1976asymptotic,
  title={Asymptotic results for goodness-of-fit statistics with unknown parameters},
  author={Stephens, Michael A},
  journal={The annals of statistics},
  pages={357--369},
  year={1976},
  publisher={JSTOR}
}

@article{solomon1978approximations,
  title={Approximations to density functions using Pearson curves},
  author={Solomon, Herbert and Stephens, Michael A},
  journal={Journal of the American Statistical Association},
  volume={73},
  number={361},
  pages={153--160},
  year={1978},
  publisher={Taylor \& Francis}
}

@misc{becker2025pearsonds,
  author       = {Becker, Martin and Klö{\ss}ner, Stefan},
  title        = {{PearsonDS}: Pearson Distribution System},
  year         = {2025},
  note         = {R package version 1.3.2},
  doi          = {10.32614/CRAN.package.PearsonDS},
  url          = {https://CRAN.R-project.org/package=PearsonDS}
}

@article{scholz1987k,
  title   = {K-Sample Anderson-Darling Tests},
  author  = {Scholz, F. W. and Stephens, M. A.},
  journal = {Journal of the American Statistical Association},
  volume  = {82},
  number  = {399},
  pages   = {918--924},
  year    = {1987}
}

@incollection{batina2022implementation,
  title={On implementation-level security of edge-based machine learning models},
  author={Batina, Lejla and Bhasin, Shivam and Breier, Jakub and Hou, Xiaolu and Jap, Dirmanto},
  booktitle={Security and Artificial Intelligence: A Crossdisciplinary Approach},
  pages={335--359},
  year={2022},
  publisher={Springer}
}

\vfill

\end{document}